\documentstyle[12pt,ere96]{article}
\begin{document}
\heading{THE STOCHASTIC GRAVITATIONAL--WAVE BACKGROUND PRODUCED BY 
NON--LINEAR COSMOLOGICAL PERTURBATIONS}

\author{Sabino Matarrese $^{1}$ and Silvia Mollerach $^{2}$} 
{$^{1}$ Dipartimento di Fisica `Galileo Galilei', Universit\`a di Padova, 
via Marzolo 8, 35131 Padova Italy.} 
{$^{2}$ Departamento de Astronom\'{\i}a y Astrof\'{\i}sica, Universidad de 
Valencia, 46100 Burjassot, Valencia, Spain.}

\begin{abstract}
\baselineskip .34cm
The cosmological stochastic gravitational-wave background produced by the 
mildly non-linear evolution of density fluctuations is analyzed, 
in the frame of an Einstein-de Sitter model, by means of a 
fully relativistic perturbation expansion up to second order. 
The form of these gravitational-instability-induced gravitational waves 
is gauge-dependent. In the synchronous gauge, where the second-order 
expansion is most easily carried out, the transverse and traceless tensor 
modes which are produced also contain a Newtonian and
post-Newtonian piece, whose interpretation as gravitational waves is 
non-trivial. 
A more direct physical understanding of this background is obtained in the
so-called Poisson gauge, where it is seen to consist of a constant term plus 
an oscillating piece whose amplitude decays inside the Hubble radius. 
\end{abstract}

\section{Introduction}

In these notes we will review some recent results concerning the generation 
of gravitational waves from the gravitational instability of 
density perturbations in a cosmological framework. 
The mixing of different modes (scalars, vectors and tensors) is a 
consequence of non-linearity, which already arises in second-order 
perturbation theory. To this aim we will first consider the exact relativistic 
dynamics of irrotational dust in the Einstein-de Sitter background
(i.e. spatially flat Robertson-Walker geometry, with vanishing 
cosmological constant), in a synchronous and comoving frame. We will then 
expand the exact equations in this gauge, up to second order in the amplitude 
of deviations from the background solution. A second-order gauge 
transformation is then performed in order to obtain the metric perturbations 
in the so-called Poisson gauge, where the tensor modes are most easily 
interpreted as gravitational radiation. 

Throughout these notes we use signature $+2$ for the metric; Greek indices
take values from 0 to 3, and Latin ones from 1 to 3.

The line-element will be written in the form 
\begin{equation}
ds^2 = a^2(\tau) g_{\mu\nu} d x^\mu d x^\nu \;, 
\end{equation}
where $a(\tau)$ is the scale-factor of the background flat Robertson-Walker 
universe and $\tau$ is the conformal time. We deal with a pressureless 
fluid, with vanishing cosmological constant, so the background 
Einstein-de Sitter model has $a(\tau) \propto \tau^2$. 
The components of the perturbed conformal metric tensor can be written as 
\begin{equation}
g_{00}= - \left[ 1 + 2\sum_{r=1}^{\infty}{1\over r!}\ \psi^{(r)}
\right] 
\;,
\end{equation}
\begin{equation}
g_{0i} = \sum_{r=1}^{\infty}{1\over r!}\ \omega^{(r)}_i\;,
\end{equation}
\begin{equation}
g_{ij} = \left[1-2\left(\sum_{r=1}^{\infty}{1\over
r!}\ \phi^{(r)}\right)\right]\delta_{ij}+\sum_{r=1}^{\infty}{1\over
r!}\ \chi^{(r)}_{ij} \;,
\end{equation}
where $\chi^{(r)i}_{i}=0$. Indices are raised and lowered by the
Kronecker symbols $\delta^{ij}$ and $\delta_{ij}$, respectively. 
The functions $\psi^{(r)}$, $\omega^{(r)}_i$, $\phi^{(r)}$, and 
$\chi^{(r)}_{ij}$ represent the $r$-th order perturbations of the metric.

We can split the perturbations into scalar (i.e. longitudinal), 
vector (i.e. transverse) and tensor (i.e. transverse and traceless) modes. 
In particular, $\omega_i^{(r)}$ can be decomposed as
\begin{equation}
\omega_i^{(r)}= \partial_i\omega^{(r)\|} +\omega_i^{(r)\perp}\;,
\end{equation}
where $\omega_i^{(r)\perp}$ is the vector component, i.e.,
$\partial^i\omega_i^{(r)\perp}=0$.  
Similarly, the traceless part of the spatial metric can be decomposed 
as 
\begin{equation}
\chi^{(r)}_{ij}= {\rm D}_{ij}\chi^{(r)\|}+\partial_i\chi^{(r)\bot}_j+
\partial_j\chi^{(r)\bot}_i+\chi^{(r)\top}_{ij}\;,
\end{equation}
where $\chi^{(r)\|}$ is a scalar potential, $\chi^{(r)\bot}_i$ are 
vector modes, and $\chi^{(r)\top}_{ij}$ tensor modes 
($\partial^i\chi^{(r)\top}_{ij}=0$); in 
the equation above we introduced the symbol 
\begin{equation}
{\rm D}_{ij} \equiv \partial_i\partial_j-
{1\over 3}\,\delta_{ij}\nabla^2\;.
\end{equation}

It is also useful to introduce similar expansions for the density 
contrast 
\begin{equation}
\delta \equiv {\varrho \over \varrho_b} -1  = 
\sum_{r=1}^{\infty}{1\over r!} ~ \delta^{(r)} \;, 
\end{equation}
with $\varrho({\bf x},\tau)$ the matter density and 
$\varrho_b(\tau)= 3/2\pi G a^2(\tau)\tau^2$ its background mean value, 
and the fluid four-velocity $u^\mu$, 
\begin{equation}
u^\mu=\frac{1}{a}\left(\delta^\mu_0 +\sum_{r=1}^{\infty}{1\over
r!}\ v^\mu_{(r)}\right) \;.
\end{equation}
The four-vector $u^\mu$ is subject to the normalization condition $u^\mu
u_\mu=-1$; therefore at any order the time component $v^0_{(r)}$ is
related to the perturbation $\psi_{(r)}$.  For the first and
second-order perturbations one obtains \cite{bmms} 
\begin{eqnarray}
\label{eq:v0psi1}
v^0_{(1)} &= &-\psi_{(1)}\; , \\
\label{eq:v0psi2}
v^0_{(2)} &= &-\psi_{(2)}+3\psi^2_{(1)}+2\omega^{(1)}_i v^i_{(1)}
+v^{(1)}_iv_{(1)}^i\; .
\end{eqnarray}

The velocity perturbation $v^i_{(r)}$ can also be split into scalar
and vector parts 
\begin{equation}
v^i_{(r)}=\partial^i v_{(r)}^{\|} +v^i_{(r)\perp}\;.
\end{equation}

\section{Relativistic dynamics of irrotational dust in Lagrangian coordinates}

We start by writing the Einstein's equations for a perfect fluid of 
irrotational dust in the synchronous ($g_{00} = -1$, $g_{0i} = 0$) and 
comoving ($u^\mu = \delta^\mu_0/a$) gauge. 
The formalism outlined in this section is discussed in greater detail in 
Ref. \cite{mt96}. 

The line--element is now written in the form 
\begin{equation}
ds^2 = a^2(\tau)\big[ - d\tau^2 + g_{ij}({\bf x}, \tau) 
dx^i dx^j \big] \;, 
\end{equation} 
with the spatial coordinates ${\bf x}$ representing Lagrangian coordinates for 
the fluid elements. 

By subtracting the isotropic Hubble--flow, one introduces the extrinsic 
curvature of constant $\tau$ hypersurfaces, 
\begin{equation}
\vartheta^i_{~j} = {1 \over 2} g^{ik}
g_{kj}' \;, 
\end{equation}
with a prime denoting differentiation with respect to the conformal time 
$\tau$. 

One can then write the Einstein's equations in a cosmologically convenient 
form. The energy constraint reads 
\begin{equation}
\vartheta^2 - \vartheta^i_{~j} \vartheta^j_{~i} + {8 \over \tau} 
\vartheta + {\cal R} = {24 \over \tau^2} \delta \;,
\end{equation}
where ${\cal R}^i_{~j}(g)$ is the intrinsic curvature
of constant time hypersurfaces, i.e. 
the conformal Ricci curvature of the three--space with 
metric $g_{ij}$ and ${\cal R}= {\cal R}^i_{~i}$. 

The momentum constraint reads 
\begin{equation}
\vartheta^i_{~j|i} = \vartheta_{,j} \;,
\end{equation}
where the vertical bar indicates a covariant derivative in the three--space 
with metric $g_{ij}$. 

Finally, after replacing the density from the energy constraint 
and subtracting the background contribution, the evolution equation for the 
extrinsic curvature reads 
\begin{equation}
{\vartheta^i_{~j}}' + {4 \over \tau} \vartheta^i_{~j} + 
\vartheta \vartheta^i_{~j} + {1 \over 4} 
\biggl( \vartheta^k_{~\ell} \vartheta^\ell_{~k} - \vartheta^2 \biggr) 
\delta^i_{~j} + {\cal R}^i_{~j} 
- {1 \over 4} {\cal R} \delta^i_{~j} 
= 0 \;. 
\end{equation}

Also useful is the Raychaudhuri equation for the evolution of the 
peculiar volume expansion scalar $\vartheta$, namely
\begin{equation}
\vartheta' + {2 \over \tau } \vartheta + \vartheta^i_{~j} \vartheta^j_{~i} 
+ {6 \over \tau^2} \delta = 0 \;. 
\end{equation}
An advantage of this gauge is that there are only geometric quantities 
in the equations, namely the spatial metric tensor with its time and space 
derivatives. The only remaining variable, the 
density contrast, can indeed be rewritten in terms of $g_{ij}$, 
by solving the continuity equation. We obtain
\begin{equation}
\delta({\bf x}, \tau) = (1 + \delta_0({\bf x})) \bigl[g({\bf x}, 
\tau)/ g_0 ({\bf x}) \bigr]^{-1/2} - 1 \;,
\end{equation}
with $g \equiv {\rm det} ~g_{ij}$. 

\section{Perturbative approach in the synchronous gauge}

In this section we will obtain the perturbations of the
Einstein-de Sitter cosmological model up to second order in the synchronous 
and comoving gauge. Different approaches to 
this problem have been used so far. 
The first solution of the second-order relativistic equations has been
obtained, in this gauge, in a pioneering work by Tomita \cite{tom1}.
Matarrese, Pantano \& Saez \cite{mps1,mps2} obtained
the leading order terms of the expansion, using a different method,
based on the so-called fluid-flow approach (e.g. Ref. \cite{ellis}). 
Salopek, Stewart and
Croudace \cite{ssc} used a gradient expansion technique to obtain
second-order metric perturbations; an intrinsic limitation of their
method is, however, that non-local terms, such as the non-linear
tensor modes, are lost. Russ {\em et al.} \cite{russ} recently
rederived the metric perturbations to second order in the synchronous
gauge, using a tetrad formalism. 

We start by expanding the covariant conformal spatial metric tensor up to 
second order in the form 
\begin{equation} 
g_{ij} = \delta_{ij} + g^{(1)}_{ij} + {1 \over 2} g^{(2)}_{ij} \;. 
\end{equation}

The contravariant metric takes the form
\begin{equation}
g^{ij} = \delta^{ij} - g^{(1)ij} - {1 \over 2} g^{(2)ij} + g^{(1)ik} 
g^{(1)j}_{k} \;.
\end{equation} 

The extrinsic curvature tensor $\vartheta^i_{~j}$ up to second order reads 
\begin{equation}
\vartheta^i_{~j} = {1 \over 2} \biggl( {g^{(1)i}_{~j}}' + {1 \over 2} 
{g^{(2)i}_{~j}}' - g^{(1)ik} {g^{(1)}_{kj}}' \biggr) \;. 
\end{equation} 

The square root of the metric determinant is 
\begin{equation}
g^{1/2} = 1 + {1 \over 2} g^{(1)i}_{~i} + {1 \over 4} g^{(2)i}_{~i} 
+ {1 \over 8} \bigl(g^{(1)i}_{~i}\bigr)^2 - {1 \over 4} g^{(1)ij} 
g^{(1)}_{ij} \;,
\end{equation} 
with inverse 
\begin{equation}
g^{-1/2} = 1 - {1 \over 2} g^{(1)i}_{~i} - {1 \over 4} g^{(2)i}_{~i} 
+ {1 \over 8} \bigl(g^{(1)i}_{~i}\bigr)^2 + {1 \over 4} g^{(1)ij} 
g^{(1)}_{ij}\;, 
\end{equation} 
so that the density contrast reads 
\begin{eqnarray}
\delta &=& - {1 \over 2} g^{(1)i}_{~i} + {1 \over 2} g^{(1)i}_{0i} 
+ \delta_0 - {1 \over 4} g^{(2)i}_{~i} + {1 \over 4} g^{(2)i}_{0i} 
+ {1 \over 8} \bigl(g^{(1)i}_{~i}\bigr)^2 + {1 \over 8} 
\bigl(g^{(1)i}_{0i}\bigr)^2 
- {1 \over 4} g^{(1)i}_{~i} g^{(1)j}_{0j} 
\nonumber\\ 
& + &
{1 \over 4} g^{(1)ij} g^{(1)}_{ij} 
- {1 \over 4} 
g^{(1)ij}_0 g^{(1)}_{0ij} - {1 \over 2} g^{(1)i}_{i} \delta_0 
+ {1 \over 2} g^{(1)i}_{0i} \delta_0 \;, 
\end{eqnarray} 
having assumed as initial conditions $g^{(2)}_{0ij} = 0$ and 
$\delta^{(2)}_0=0$ (i.e. $\delta_0=\delta^{(1)}_0$). 

The Christoffel symbols up to second order read 
\begin{equation}
\Gamma^i_{jk} = {1 \over 2} \biggl( g^{(1)i}_{~j~,k} + g^{(1)i}_{~k~,j} -
g^{(1),i}_{jk} \biggr) + {1 \over 4} 
\biggl( g^{(2)i}_{~j~,k} + g^{(2)i}_{~k~,j} - g^{(2),i}_{jk} \biggr) - 
{1 \over 2} g^{(1)i}_{~\ell} \biggl( g^{(1)\ell}_{~j~,k} + g^{(1)\ell}_{~k~,j} 
- g^{(1),\ell}_{jk} \biggr) \;, 
\end{equation} 
from which, after a lenghty but straightforward calculation, the conformal 
Ricci tensor of the spatial hypersurfaces is obtained
\begin{eqnarray}
{\cal R}^i_{~j} & = & {1 \over 2} \biggl( g^{(1)ik}_{~~~~,jk} + 
g^{(1)k~,i}_{~j~,k} - \nabla^2 g^{(1)i}_{~j} - g^{(1)k~,i}_{~k~,j} 
\biggr) + {1 \over 4} \biggl( g^{(2)ik}_{~~~~,jk} + 
g^{(2)k~,i}_{~j~,k} - \nabla^2 g^{(2)i}_{~j} - g^{(2)k~,i}_{~k~,j} \biggr) 
\nonumber\\
& + & {1 \over 2} \biggl[ g^{(1)ik} \biggl( \nabla^2 g^{(1)}_{kj} 
+ g^{(1)\ell}_{~\ell~,jk} - g^{(1)\ell}_{~k~,\ell j} - 
g^{(1)\ell}_{~j~,\ell k} \biggr) + 
g^{(1)\ell k} \biggl( g^{(1)i}_{~j~,k\ell} + g^{(1),i}_{~\ell k~,j} -
g^{(1)i}_{~k~,j\ell} 
\nonumber\\
& - & g^{(1),i}_{~j\ell~,k} \biggr) 
+ g^{(1)\ell k}_{~~~~~,\ell} \biggl( g^{(1)i}_{~j~,k} 
+ g^{(1)i}_{~k~,j} - g^{(1),i}_{jk} \biggr) 
+ g^{(1)\ell i}_{~~~~,k} g^{(1),k}_{~\ell j} 
- g^{(1)\ell i}_{~~~~,k} g^{(1)k}_{~j~,\ell} 
\nonumber\\
& + & {1 \over 2} g^{(1)\ell k}_{~~~~,j} g^{(1),i}_{~\ell k} 
+ {1 \over 2} g^{(1)m}_{~m~,\ell} \biggl( g^{(1)\ell i}_{~~~~,j} + 
g^{(1)\ell,i}_{~j} - g^{(1)i,\ell}_{~j} \biggr) \biggr] 
\; ; 
\end{eqnarray} 
its trace reads
\begin{eqnarray}
{\cal R}  & = & g^{(1)\ell k}_{~~~~~,\ell k} - g^{(1)k~,\ell}_{~k~~,\ell} 
+ {1 \over 2} \biggl( g^{(2)\ell k}_{~~~~~,\ell k} - 
g^{(2)k~,\ell}_{~k~~,\ell} \biggr) + g^{(1)jk} \biggl( \nabla^2 g^{(1)}_{jk} 
+ g^{(1)\ell}_{~\ell~,jk} - 2 g^{(1)\ell}_{~j~,\ell k} \biggr) 
\nonumber\\
& + & 
 g^{(1)\ell k}_{~~~~,\ell} \biggl( g^{(1)j}_{~j~,k} - g^{(1),j}_{jk} \biggr) 
+ {3 \over 4} g^{(1)\ell j}_{~~~~,k} g^{(1),k}_{~\ell j} 
- {1 \over 2} g^{(1)\ell j}_{~~~~,k} g^{(1)k}_{~j~,\ell} 
- {1 \over 4} g^{(1)j,\ell}_{~j} g^{(1)k}_{~k~,\ell} 
\;. 
\end{eqnarray} 

\subsection{First-order perturbations} 

We are now ready to deal with the equations above at the linear level. 
Let us then write the conformal spatial metric tensor in the form 
\begin{equation}
g_{ij} = \delta_{ij} + g^{(1)}_{Sij} \;, 
\end{equation}
where, from now on, we use the subscript $S$ to label synchronous-gauge 
perturbations. 

We then write
\begin{equation}
g^{(1)}_{Sij} = - 2 \phi_S^{(1)} \delta_{ij} + 
{\rm D}_{ij}\chi^{(1)\|}_S +\partial_i\chi^{(1)\bot}_{Sj} +
\partial_j\chi^{(1)\bot}_{Si}+ \chi^{(1)\top}_{ij}\;, 
\end{equation}
with
\begin{equation}
\partial^i\chi^{(1)\bot}_{Si}= \chi^{(1)\top i}_{~~i} = 
\partial^i \chi^{(1)\top}_{ij} = 0 \;,
\end{equation}
indices being raised by $\delta^{ij}$. Remind that at first order 
the tensor modes $\chi^{(1)\top}_{ij}$ are gauge-invariant. 

As well known, in linear theory, scalar, vector and tensor modes are 
independent. The equation of motion for the tensor modes 
is obtained by linearizing the traceless part of the 
$\vartheta^i_{~j}$ evolution equation. One has
\begin{equation}
{\chi^{(1)\top}_{ij}}'' +  {4 \over \tau} {\chi^{(1)\top}_{ij}}' 
- \nabla^2 \chi^{(1)\top}_{ij} = 0 \;,
\end{equation}
which is the equation for the free propagation of gravitational 
waves in the Einstein-de Sitter universe. The general solution of this 
equation is 
\begin{equation}
\chi^{\top(1)}_{ij}({\bf x},\tau)=\frac{1}{(2\pi)^3} \int d^3{\bf k}
\exp(i{\bf k}\cdot{\bf x}) \chi^{(1)}_\sigma({\bf k},\tau)
\epsilon^{\sigma}_{ij}(\hat{\bf k}),
\end{equation}
where $\epsilon^{\sigma}_{ij}(\hat{\bf k})$ is the polarization tensor,
with $\sigma$ ranging over the polarization components $+,\times$, and
$\chi^{(1)}_\sigma({\bf k},\tau)$ are the amplitudes of the two polarizationm 
states, whose time evolution can be represented as
\begin{equation}
\chi^{(1)}_\sigma({\bf k},\tau) = A(k) a_\sigma({\bf k})
\left(\frac{3 j_1(k\tau)}{k\tau}\right),
\end{equation}
with $j_1$ the spherical Bessel function of order 1 and $a_\sigma({\bf k})$ a 
zero mean random variable with auto-correlation function
\begin{equation}
\langle a_\sigma({\bf k}) a_{\sigma'}({\bf k'})\rangle =(2\pi)^3
k^{-3} \delta^3({\bf k} + {\bf k'}) \delta_{\sigma\sigma'}\;. 
\end{equation}
The spectrum of the gravitational-wave background depends on the
processes by which it was generated, and for example in most
inflationary models, $A(k)$ is nearly scale invariant and proportional
to the Hubble constant during inflation.
 
In the irrotational case the linear vector perturbations 
represent gauge modes which can be set to zero: $\chi^{(1)\bot}_i = 0$. 

The two scalar modes are linked together via the momentum constraint, 
leading to the condition 
\begin{equation}
\phi_S^{(1)} + {1\over 6} \nabla^2 \chi^{(1)\|}_S = 
\phi^{(1)}_{S0} + {1\over 6} \nabla^2 \chi^{(1)\|}_{S0} \;. 
\end{equation}

The energy constraint gives
\begin{equation}
\nabla^2 \biggl[ {2 \over \tau} {\chi^{(1)\|}_S}' + 
{6 \over \tau^2} \bigl( \chi^{(1)\|}_S - \chi^{(1)\|}_{S0} \bigr) 
+ 2 \phi^{(1)}_{S0} + {1 \over 3} \nabla^2 \chi^{(1)\|}_S \biggr] = 
{12 \over \tau^2} \delta_0 \;, 
\end{equation}
having consistently assumed $\delta_0\ll 1$. 

The trace of the evolution equation also gives an equation for the scalar 
modes, 
\begin{equation}
{\chi_S^{(1)\|}}''+ {4 \over \tau} {\chi_S^{(1)\|}}' + {1 \over 3} 
\nabla^2 \chi^{(1)\|}_S = - 2 \phi_S^{(1)} \;.
\end{equation}
An equation only for the scalar mode $\chi_S^{(1)\|}$ can be obtained by 
combining together the evolution equation and the energy constraint,
\begin{equation}
\nabla^2 \biggl[ {\chi_S^{(1)\|}}'' + {2 \over \tau} {\chi_S^{(1)\|}}' 
- {6 \over \tau^2} \bigl( \chi^{(1)\|}_S - \chi^{(1)\|}_{S0} \bigr) \biggr] = - 
{2 \over \tau^2} \delta_0 \;. 
\end{equation}

On the other hand, by linearizing the solution of the continuity 
equation, we obtain 
\begin{equation}
\delta_S^{(1)} = \delta_0 - {1 \over 2} 
\nabla^2 \bigl(\chi^{(1)\|}_S - \chi^{(1)\|}_{S0} \bigr) \;,
\end{equation}
which replaced in the previous expression gives 
\begin{equation}
{\delta^{(1)}_S}'' + {a' \over a} {\delta_S^{(1)}}' - 4\pi G a^2 
\varrho_b \delta_S^{(1)} = 0 \;.
\end{equation}
This is the equation for linear density fluctuations 
(e.g. Ref. \cite{pjep80}) whose general solution is 
straightforward to obtain. 

The equations above have been obtained in whole generality; 
one could have used 
instead the well-known residual gauge ambiguity of the synchronous 
coordinates (e.g. Refs. \cite{mt96,russ}) to simplify 
their form. 
For instance, one could fix $\chi^{(1)\|}_0$ so that 
$\nabla^2 \chi^{(1)\|}_{S0} = - 2 \delta_0$, so that the 
$\chi^{(1)\|}_S$ evolution equation takes the same 
form as that for $\delta_S^{(1)}$. With such a gauge fixing 
one obtains 
\begin{equation}
\chi^{(1)\|}_S({\bf x},\tau) = \chi_+({\bf x}) \tau^2 + 
\chi_-({\bf x}) \tau^{-3} \;,
\end{equation} 
where $\chi_\pm$ set the amplitudes of the growing ($+$) and 
decaying ($-$) modes. 
In what follows, we shall restrict ourselves to the growing 
mode. The effect of the decaying mode on second-order perturbations has 
been considered by Tomita \cite{tom1} and by Russ et al. \cite{russ} 
and will not be studied here. 
The amplitude of the growing mode is related to the initial {\em peculiar 
gravitational potential}, through $\chi_+ \equiv - {1 \over 3} 
\varphi$, where in turn, $\varphi$ is related to $\delta_0$ through the 
cosmological Poisson equation 
\begin{equation}
\nabla^2 \varphi({\bf x}) = {6 \over \tau_0^2} 
\delta_0({\bf x}) \;. 
\end{equation}
Therefore, 
\begin{equation}
{\rm D}_{ij}\chi^{(1)\|}_S = 
- {\tau^2 \over 3} \biggl( \varphi_{,ij} 
- {1 \over 3} \delta_{ij} \nabla^2 \varphi \biggr) \;.
\end{equation}
The remaining scalar mode reads
\begin{equation}
\phi_S^{(1)}({\bf x},\tau) = {5 \over 3} \varphi({\bf x}) + {\tau^2 \over 18} 
\nabla^2 \varphi({\bf x}) \;. 
\end{equation}

In what follows we will also drop the tensor mode $\chi^{(1)\top}_{ij}$ 
as a possible source for second-order perturbations, i.e. we will restrict 
ourselves to initial growing-mode scalar perturbations. The effect 
of these modes on second-order perturbations has been studied by Tomita 
\cite{tom2,tom3} and Matarrese, Mollerach \& Bruni \cite{mmb}.

\subsection{Second-order perturbations} 

Let us start by the definition
\begin{equation} 
g^{(2)}_{Sij} = - 2 \phi_S^{(2)} \delta_{ij} + \chi^{(2)}_{Sij} \;,
\end{equation}
with $\chi^{(2)i}_{S~~~j}=0$. 

The technique of second-order perturbation theory is straightforward: with the 
help of the relations reported at the beginning of Section 3
[Eqs.(20) - (28)], we first 
substitute the expansion above in our exact fluid-dynamical equations 
(momentum and energy constraints plus evolution and Raychaudhuri 
equations) obtaining equations for $g^{(2)}_{Sij}$, with source terms 
containing quadratic combinations of $g^{(1)}_{Sij}$ plus a few more terms 
involving $\delta_0$. Next, we have to solve these equations for the 
modes $\phi_S^{(2)}$ and $\chi^{(2)}_{Sij}$ in terms of the initial 
peculiar gravitational potential $\varphi$.

Let us now give the equations which govern the evolution of the second-order 
metric perturbations. 

The second-order Raychaudhuri equation reads

\begin{eqnarray} 
{\phi_S^{(2)}}'' & + & {2 \over \tau} {\phi_S^{(2)}}' - {6 \over \tau^2} 
\phi_S^{(2)} = - {1 \over 6} {g^{(1)ij}_S}' \biggl( 
{g^{(1)}_{Sij}}' - { 4 \over \tau} g^{(1)}_{Sij} \biggr) 
+ {1 \over 6} \biggl[ 2 g^{(1)ij}_S \biggl( 2 g^{(1)k}_{Si~,kj} 
- \nabla^2 g^{(1)}_{Sij} - g^{(1)k}_{Sk~,ij} \biggr) 
\nonumber\\
& - &  g^{(1)k}_{Sk} \biggl(g^{(1)ij}_{S~~~,ij} - \nabla^2 g^{(1)i}_{Si} 
\biggr) 
\biggr] - {2 \over \tau^2} \biggl[ - {1 \over 4} \biggl(g^{(1)i}_{Si} 
- g^{(1)~i}_{S0i} \biggr)^2 - {1 \over 2} 
\biggl(g^{(1)ij}_S g^{(1)}_{Sij} - g^{(1)ij}_{S0} g^{(1)}_{S0ij} \biggr) 
\nonumber\\
& + & \delta_0 \biggl(g^{(1)i}_{Si} - g^{(1)~i}_{S0i} \biggr) \biggr] 
\;. 
\end{eqnarray} 

The second-order energy constraint reads

\begin{eqnarray}
{2 \over \tau} {\phi^{(2)}_S}' & - & {1 \over 3} 
\nabla^2 \phi^{(2)}_{S} +{6 \over \tau^2} 
\phi^{(2)}_{S} - {1 \over 12} \chi^{(2)ij}_{S~~~,ij} 
= - {2 \over 3 \tau} g^{(1)ij}_S {g^{(1)}_{Sij}}' - 
{1 \over 24} \biggl( {g^{(1)ij}_S}' {g^{(1)}_{Sij}}' - {g^{(1)i}_{Si}}'
{g^{(1)j}_{Sj}}'\biggr) 
\nonumber\\ 
& + & 
{1 \over 6} \biggl[ g^{(1)ij}_S \biggl( \nabla^2 g^{(1)}_{Sij}
+ g^{(1)k}_{Sk~,ij} - 2 g^{(1)k}_{Si~,jk} \biggr) + 
g^{(1)ki}_{S~~~,k} \biggl( g^{(1)j}_{Sj~,i} - g^{(1)j}_{Si~,j} \biggr) 
+ {3 \over 4} g^{(1)ij}_{S~~~,k} g^{(1),k}_{Sij} 
\nonumber\\ 
& - & 
{1 \over 2} g^{(1)ij}_{S~~~,k} g^{(1)k}_{Si~~,j}  
- {1 \over 4} g^{(1)i,k}_{Si} g^{(1)j}_{Sj~,k} \biggr] 
+ {2 \over \tau^2} \biggl[ - {1 \over 4} \biggl(g^{(1)i}_{Si} 
- g^{(1)~i}_{S0i} \biggr)^2 - {1 \over 2} 
\biggl( g^{(1)ij}_S g^{(1)}_{Sij} 
\nonumber\\ 
& - & g^{(1)ij}_{S0} g^{(1)}_{S0ij} \biggr) 
+ \delta_0 \biggl(g^{(1)i}_{Si} - g^{(1)~i}_{S0i} \biggr) \biggr] 
\;. 
\end{eqnarray} 

The second-order momentum constraint reads 

\begin{equation}
2 {\phi^{(2)}_{S,j}}' + {1 \over 2} 
{\chi^{(2)i}_{Sj~,i}}' = g^{(1)ik}_{S} 
\biggl( {g^{(1)}_{Sjk,i}}' - {g^{(1)}_{Sik,j}}' \biggr) 
+ g^{(1)ik}_{S~~~~,i} {g^{(1)}_{Sjk}}' 
- {1 \over 2} g^{(1)ik}_{S~~~~,j} {g^{(1)}_{Sik}}' 
- {1 \over 2} g^{(1)i}_{Si~,k} {g^{(1)k}_{Sj}}' 
\;. 
\end{equation} 

Finally, the second-order evolution equation reads

\begin{eqnarray}
& - & \biggl({\phi^{(2)}_S}'' + {4 \over \tau} 
{\phi^{(2)}_S}' \biggr) \delta^i_{~j} + 
{1 \over 2} \biggl( {\chi^{(2)i}_{Sj}}'' + 
{4 \over \tau} {\chi^{(2)i}_{Sj}}' \biggr) + 
\phi^{(2),i}_{S,j} - {1 \over 4} 
\chi^{(2)k\ell}_{S~~~,k\ell} \delta^i_{~j} + 
{1 \over 2} \chi^{(2)ki}_{S~~~,kj} 
\nonumber\\ 
& + & {1 \over 2} \chi^{(2)k,i}_{Sj~~~,k} - 
{1 \over 2} \nabla^2 \chi^{(2)i}_{Sj} 
= {g^{(1)ik}_S}' {g^{(1)}_{Skj}}' - {1 \over 2} {g^{(1)k}_{Sk}}' 
{g^{(1)i}_{Sj}}' + {1 \over 8} \biggl[\biggl({g^{(1)k}_{Sk}}'\biggr)^2 - 
{g^{(1)k}_{S\ell}}' {g^{(1)\ell}_{Sk}}' \biggr] \delta^i_{~j} 
\nonumber\\ 
& - & {1 \over 2} \biggl[ - g^{(1)i}_{Sj} \biggl( g^{(1)k,\ell}_{S\ell~~~,k}
- \nabla^2 g^{(1)k}_{Sk} \biggr) 
+ 2 g^{(1)k\ell}_S \biggl( g^{(1)i}_{Sj~,k\ell} + g^{(1)~,i}_{Sk\ell~,j} 
- g^{(1)i}_{S\ell~,jk}  - g^{(1)~,i}_{S\ell j~,k}  \biggr) 
\nonumber\\ 
& + & 2 g^{(1)k\ell}_{S~~~,k} \biggl( g^{(1)i}_{Sj~,\ell} 
- g^{(1)i}_{S\ell~,j} - g^{(1),i}_{Sj\ell} \biggr) 
+ 2 g^{(1)ki}_{S~~~,\ell} g^{(1),\ell}_{Sjk} 
- 2 g^{(1)ki}_{S~~~,\ell} g^{(1)\ell}_{Sj~,k} 
+ g^{(1)k\ell}_{S~~~,j} g^{(1)~,i}_{Sk\ell} 
\nonumber\\ 
& + &
g^{(1)\ell}_{S\ell~,k} \biggl( g^{(1)ki}_{S~~~,j} + g^{(1)k,i}_{Sj} 
- g^{(1)i,k}_{Sj} \biggr) 
- g^{(1)k\ell}_S \biggl( \nabla^2 g^{(1)}_{Sk\ell} 
+ g^{(1)m}_{Sm~,k\ell} - 2 g^{(1)m}_{Sk~~,m\ell} \biggr) \delta^i_{~j}  
\nonumber\\ 
& - & g^{(1)\ell k}_{S~~~,\ell} \biggl( g^{(1)m}_{Sm~,k} - g^{(1)m}_{Sk~,m} 
\biggr) \delta^i_{~j} - {3 \over 4} g^{(1)k\ell}_{S~~~,m} g^{(1),m }_{Sk\ell} 
\delta^i_{~j} + {1 \over 2} g^{(1)k\ell}_{S~~~,m} g^{(1)m }_{Sk~~,\ell} 
\delta^i_{~j} 
\nonumber\\ 
& + & {1 \over 4} g^{(1)k,m}_{Sk} g^{(1)\ell}_{S\ell~,m} 
\delta^i_{~j} 
\;. 
\end{eqnarray} 

The next step is to solve these equations. 
In these calculations, we can make the simplifying assumption that the initial 
conditions are taken at conformal time $\tau_0=0$. 
One can start from the Raychaudhuri equation, to obtain the trace of the 
second-order metric tensor. (Actually, in order to obtain the sub-leading term
we also need to use the energy constraint). The 
resulting expression for $\phi_{S}^{(2)}$ is 
\begin{equation}
\phi_{S}^{(2)}=\frac{\tau^4}{252}\left(-\frac{10}{3}\varphi^{,ki}
\varphi_{,ki}+(\nabla^2\varphi)^2\right)
+{5 \tau^2 \over 18}\left(\varphi^{,k}\varphi_{,k}+
\frac{4}{3}\varphi\nabla^2\varphi\right) 
\;. 
\end{equation}

The expression for $\chi^{(2)}_{Sij}$ is obtained by first replacing
$\phi_{S}^{(2)}$ into the remaining equations and solving first 
the energy constraint, next the momentum constraint and finally 
the (traceless part of the) evolution equation. 
We obtain 
\begin{eqnarray}
\chi_{Sij}^{(2)}& = &\frac{\tau^4}{126}\left(19\varphi^{,k}_{~~,i}
\varphi_{,kj}-12 \varphi_{,ij} \nabla^2\varphi
+4 (\nabla^2\varphi)^2 \delta_{ij}
-\frac{19}{3}\varphi^{,kl}\varphi_{,kl} \delta_{ij}\right)\nonumber\\
&+&\frac{5\tau^2}{9}\left(-6\varphi_{,i}\varphi_{,j}
-4\varphi \varphi_{,ij}+2 \varphi^{,k}\varphi_{,k}\delta_{ij}
+\frac{4}{3}\varphi\nabla^2\varphi\delta_{ij}\right)
+\pi_{ij} \;. 
\end{eqnarray}
The transverse and traceless contribution $\pi_{ij}$, which 
represents the second-order tensor mode $\chi^{(2)\top}_{Sij}$ generated by 
(growing-mode) scalar initial perturbations, is determined by the 
inhomogeneous wave-equation 
\begin{equation}
\pi_{ij}''+\frac{4}{\tau}\pi_{ij}'-\nabla^2 \pi_{ij}=
-\frac{\tau^4}{21}\nabla^2 {\cal S}_{ij}\;,
\end{equation}
with
\begin{equation}
{\cal S}_{ij}=\nabla^2 \Psi_0 \delta_{ij}+ \Psi_{0,ij}+
2\left(\varphi_{,ij}\nabla^2\varphi-\varphi_{,ik}
\varphi^{,k}_{~~,j}\right)\;,
\end{equation}
where
\begin{equation}
\nabla^2 \Psi_0=-\frac{1}{2}\left((\nabla^2\varphi)^2-
\varphi_{,ik}\varphi^{,ik}\right)\;.
\end{equation}
This equation can be solved by the Green's method; we obtain for 
$\pi_{ij}$ that
\begin{equation}
\pi_{ij}({\bf x},\tau) = \frac{\tau^4}{21}{\cal S}_{ij}({\bf x}) + 
\frac{4\tau^2}{3} {\cal T}_{ij}({\bf x})
+\tilde{\pi}_{ij}({\bf x},\tau)\;,
\end{equation}
where $\nabla^2 {\cal T}_{ij} \equiv {\cal S}_{ij}$ and the remaining piece
$\tilde{\pi}_{ij}$, containing a term that is
constant in time and another one that oscillates with decreasing
amplitude just like the linear tensor modes, can be written as
\begin{equation}
\tilde{\pi}_{ij}({\bf x},\tau)=\frac{1}{(2\pi)^3} \int d^3{\bf k}
\exp(i{\bf k}\cdot{\bf x})\frac{40}{k^4} {\cal S}_{ij}({\bf k})\left(
\frac{1}{3} - \frac{j_1(k\tau)}{k\tau} \right)\;,
\end{equation}
with ${\cal S}_{ij}({\bf k})=\int d^3{\bf x}\exp(-i{\bf k}\cdot{\bf x})
{\cal S}_{ij}({\bf x})$.

\section{Perturbative approach in the Poisson gauge}

The Poisson gauge, recently discussed by Bertschinger
\cite{bert}, is uniquely defined by 
${\omega_{Pi}}^{(r),i}={\chi_{Pij}}^{(r),j}=0$. 
This gauge generalizes the so-called longitudinal gauge to include vector 
and tensor modes.  The latter gauge, in which $\omega_{Li}^{(r)}=
\chi_{Lij}^{(r)}=0$, has 
been widely used in the literature to investigate the linear evolution of
scalar perturbations \cite{mfb}. Since vector and tensor modes
are set to zero by hand, the longitudinal gauge cannot be used to
study perturbations beyond the linear regime, because in the nonlinear
case vector and tensor modes are dynamically generated by the 
growth of scalar modes. 

Instead of writing the perturbed Einstein's equations directly in this gauge, 
we will transform the synchronous gauge quantities, by means of the 
second-order gauge transformations introduced in \cite{bmms}. 
This problem has been dealt with in detail in Refs. \cite{bmms,mmb}. 

\subsection{Transforming from the synchronous to the Poisson gauge}

In these notes, we are only interested in calculating first and 
second-order perturbations of the metric tensor. A 
gauge transformation of order $r$ can be associated to a  
coordinate transformation
along $r$ independent vector fields $\xi_{(r)}^\nu$, given by 
\begin{equation}
\tilde{x}^\mu =x^\mu - \xi_{(1)}^\mu +
\frac{1}{2}\left(\xi_{(1),\nu}^\mu \xi_{(1)}^\nu -
\xi_{(2)}^\mu\right)+ \cdots \; .
\end{equation}
The general gauge transformation rule for a tensor $T$, with background 
value $T_0$ and fluctuation $\Delta T = \sum_{r=1}^\infty (1/r!) \delta^r T$ is
\begin{equation}
\delta \tilde{T} =\delta T +\pounds_{\xi_{(1)}} T_{0}\; ,
\end{equation}
to first order, 
\begin{equation}
\delta^2 \tilde{T} =\delta^2 T +2\pounds_{\xi_{(1)}} \delta T
+\pounds^{2}_{\xi_{(1)}} T_{0} +\pounds_{\xi_{(2)}}
T_{0}\;, 
\end{equation}
to second order, etc.. Here $\pounds_\xi$ is the first-order Lie derivative 
along the vector $\xi^\mu$, $\pounds^2_\xi$ is the second-order one, etc..,
and we remind that the Lie derivative acting on a covariant tensor 
$T_{\mu\nu}$ of rank two (which is the case we are interested in here) is 
\begin{equation}
\pounds_\xi T_{\mu\nu} = T_{\mu\nu,\sigma}\xi^\sigma
+ \xi^\sigma_{~~,\mu} T_{\sigma\nu} + \xi^\sigma_{~~,\nu} T_{\mu\sigma}\; .
\end{equation}
A general $r$-th order gauge transformation is then determined by a set of 
$r$ vectors $\xi_{(r)}^\nu$, whose components we can also split in 
scalar and vector modes, as usual 
\begin{equation}
\xi_{(r)}^0=\alpha^{(r)}\;,
\end{equation}
and
\begin{equation}
\xi_{(r)}^i=\partial^i\beta^{(r)}+d^{(r)i}\;,
\end{equation}
with $\partial_i d^{(r)i}=0$.

We are here only interested in performing the gauge tranformation on the 
covariant metric tensor $a^2(\tau) g_{\mu\nu}$. In transforming from 
the synchronous to the Poisson gauge, to first order, one obtains 
\cite{bmms} 
\begin{equation}
\psi_{\scriptscriptstyle \rm P}^{(1)}
=\alpha^{(1)\prime}+{2 \over \tau}\,\alpha^{(1)}\;,
\label{i}
\end{equation}
\begin{equation}
\alpha^{(1)}=\beta^{(1)\prime}\;,
\label{iii}
\end{equation}
\begin{equation}
\omega_{{\scriptscriptstyle\rm P}\ i}^{(1)}=d^{(1)\prime}_i\;,
\label{iv}
\end{equation}
\begin{equation}
\phi_{\scriptscriptstyle \rm P}^{(1)}=\phi_{\scriptscriptstyle
\rm S}^{(1)}-{1\over 3}\,\nabla^2\beta^{(1)}-
{2 \over \tau}\,\alpha^{(1)}\;, 
\label{ii}
\end{equation}
\begin{equation}
{\rm D}_{ij}\left(\chi_{\scriptscriptstyle
\rm S}^{(1)\|}+2\beta^{(1)}\right)=0\;,
\label{v}
\end{equation}
\begin{equation}
\chi_{{\scriptscriptstyle
\rm S}\ (i,j)}^{(1)\bot}+d^{(1)}_{(i,j)}=0\;,
\label{vi}
\end{equation}
\begin{equation}
\chi_{{\scriptscriptstyle\rm  P}\ ij}^{(1)\top}=
\chi_{{\scriptscriptstyle
\rm S}\ ij}^{(1)\top}\;.
\label{vii}
\end{equation}
From these equations, given the first order perturbed metric
in the synchronous gauge, we can obtain the parameters of the 
gauge transformation and the perturbed metric in the Poisson gauge.

The second-order transformation, in the particular case where only 
scalar modes are present at first order 
($\chi_{{\scriptscriptstyle \rm S}\
ij}^{(1)\bot}= \chi_{{\scriptscriptstyle \rm S}\
ij}^{(1)\top}=v^i_{(1)\perp}=0$, implying 
$d_i^{(1)}=\omega^{(1)}_{{\scriptscriptstyle \rm P}\ i}
=\chi_{{\scriptscriptstyle \rm P}\ ij}^{(1)\bot}=0$), yields 
\begin{equation}
\label{eq:psiP}
\psi^{(2)}_{\scriptscriptstyle \rm P}=\beta_{(1)}^{\prime}\left[
 \beta_{(1)}^{\prime\prime\prime}
+\frac{10}{\tau}\beta_{(1)}^{\prime\prime}
+ \frac{6}{\tau^2}
\beta_{(1)}^{\prime}\right]
+\beta_{(1)}^{ ,i}\left(
\beta^{(1)\prime\prime}_{,i}
+\frac{2}{\tau}\beta^{(1)\prime}_{,i}\right)
+2\beta_{(1)}^{\prime\prime 2}
+\alpha^{(2)\prime}+\frac{2}{\tau}\alpha^{(2)}\;,
\end{equation}
\begin{equation}\label{eq:omegaP}
\omega^{(2)}_{{\scriptscriptstyle \rm P}\ i}=
-2\left(
2\phi^{(1)}_{\scriptscriptstyle \rm S}
+\beta_{(1)}^{\prime\prime}
-\frac{2}{3}\nabla^2\beta_{(1)}\right)\beta^{(1)\prime}_{,i}
-2\beta^{(1)\prime}_{,j}\beta^{(1),j}_{,i}-
\alpha^{(2)}_{,i} +\beta^{(2)\prime}_{,i} +d^{(2) \prime}_i\;,
\end{equation}
\begin{eqnarray}\label{eq:phiP}
\phi^{(2)}_{\scriptscriptstyle \rm P}& =
 & \phi^{(2)}_{\scriptscriptstyle
\rm S}
+\beta^{\prime}_{(1)}\left[
2\left(\phi_{\scriptscriptstyle
\rm S}^{(1)\prime}+
\frac{4}{\tau}\phi^{(1)}_{\scriptscriptstyle
\rm S}\right)
- \frac{6}{\tau^2} \beta^{\prime}_{(1)}
-\frac{2}{\tau} \beta_{(1)}^{\prime\prime}\right]
\nonumber \\
& - & \frac{1}{3}\left(-4\phi^{(1)}_{\scriptscriptstyle
\rm S}+\beta_{(1)}^{\prime}\partial_0
+\beta^{,i}_{(1)}\partial_i
+\frac{8}{\tau}\beta^{\prime}_{(1)}
+\frac{4}{3}\nabla^2\beta_{(1)}\right)\nabla^2\beta_{(1)}
\nonumber \\
& + & \beta^{,i}_{(1)}\left( 2\phi^{(1)}_{{\scriptscriptstyle
\rm S} ,i}-\frac{2}{\tau}\beta^{(1)\prime}_{,i}\right)
+\frac{2}{3}\beta^{(1)}_{,ij}\beta^{,ij}_{(1)}
-\frac{2}{\tau}\alpha_{(2)} -\frac{1}{3}\nabla^2\beta_{(2)}\;,
\end{eqnarray}
\begin{eqnarray}
\chi^{(2)}_{{\scriptscriptstyle \rm P}\ ij} & = &
\chi^{(2)}_{{\scriptscriptstyle \rm S}\ ij}
+2\left(\frac{4}{3}\nabla^2\beta_{(1)}
-4\phi^{(1)}_{\scriptscriptstyle
\rm S} -
\beta^{\prime}_{(1)}\partial_0-\beta_{(1)}^{,k}\partial_k\right){\rm
D}_{ij} \beta_{(1)}
\nonumber \\
& - & 4\left(\beta^{(1)}_{,ik}\beta^{,k}_{(1),j}
-\frac{1}{3}\delta_{ij}\beta^{(1)}_{,lk}\beta_{(1)}^{,lk}\right)
+2\left(d^{(2)}_{(i,j)} +{\rm D}_{ij}\beta^{(2)}\right)\; .
\end{eqnarray}

One can, at least implicitly, compute the parameters involved in this 
second-order transformation in terms of the second order perturbed metric
in the synchronous gauge and first order perturbed quantities, namely 
\begin{eqnarray}\label{beta2}
\nabla^2\nabla^2\beta_{(2)} & = & -\frac{3}{4}
\chi^{(2),ij}_{{\scriptscriptstyle \rm S}\ ij}+
6\phi_{\scriptscriptstyle \rm S}^{(1),ij}
\beta^{(1)}_{,ij} -2 \nabla^2\phi_{\scriptscriptstyle \rm S}^{(1)}
\nabla^2\beta_{(1)}
+8  \phi_{\scriptscriptstyle \rm S}^{(1),i} \nabla^2\beta^{(1)}_{,i}
+4 \phi_{\scriptscriptstyle \rm S}^{(1)} \nabla^2\nabla^2\beta_{(1)}
 \nonumber \\
& + &
4\nabla^2\beta^{(1)}_{,ij}\beta_{(1)}^{,ij}
-\frac{1}{6}\nabla^2\beta_{(1)}^{,i} \nabla^2\beta^{(1)}_{,i}
+\frac{5}{2}\beta_{(1)}^{,ijk} \beta^{(1)}_{,ijk}
-\frac{2}{3}\nabla^2\beta_{(1)} \nabla^2\nabla^2\beta_{(1)}
+ \frac{3}{2}\beta_{(1)}^{,ij\prime} \beta^{(1)\prime}_{,ij}
\nonumber \\
& - &
\frac{1}{2}\nabla^2\beta_{(1)}^{\prime} \nabla^2\beta_{(1)}^{\prime}
+2\beta_{(1)}^{,i\prime} \nabla^2\beta^{(1)\prime}_{,i}
+\beta_{(1)}^{\prime} \nabla^2\nabla^2\beta_{(1)}^{\prime}
+\beta_{(1)}^{,i} \nabla^2\nabla^2\beta^{(1)}_{,i}\;,
\end{eqnarray}

\begin{eqnarray}
\nabla^2 d^{(2)}_i & = &-\frac{4}{3}\nabla^2\beta^{(2)}_{,i}
-\chi^{(2),j}_{{\scriptscriptstyle \rm S}\ ij}+
8\phi_{\scriptscriptstyle \rm S}^{(1),j}{\rm D}_{ij}\beta_{(1)}
+\frac{16}{3}\phi_{\scriptscriptstyle \rm S}^{(1)}
\nabla^2\beta^{(1)}_{,i}
+\frac{2}{3}\nabla^2\beta_{(1)}^{,j}\beta^{(1)}_{,ij}
+\frac{10}{3}\beta_{(1)}^{,jk}\beta^{(1)}_{,ijk}
 \nonumber \\
&- &
\frac{8}{9}\nabla^2\beta_{(1)} \nabla^2\beta^{(1)}_{,i}
+2\beta_{(1)}^{,j\prime}{\rm D}_{ij}\beta_{(1)}^{\prime}
+\frac{4}{3}\beta_{(1)}^{\prime}\nabla^2\beta^{(1)\prime}_{,i}
+\frac{4}{3}\beta_{(1)}^{,j}\nabla^2\beta^{(1)}_{,ij} \;,
\end{eqnarray}

\begin{eqnarray}
\nabla^2\alpha_{(2)} & = & \nabla^2\beta_{(2)}^{\prime}
-2\left(2\phi^{(1),i}_{\scriptscriptstyle \rm S}
+\beta_{(1)}^{\prime\prime,i} +\frac{1}{3}\nabla^2\beta_{(1)}^{,i}
\right)\beta^{(1)\prime}_{,i}
\nonumber \\
& - &
2\left(2\phi^{(1)}_{\scriptscriptstyle \rm S}
+\beta_{(1)}^{\prime\prime} -\frac{2}{3}\nabla^2\beta_{(1)}
\right)\nabla^2\beta_{(1)}^{\prime}
-2\beta_{(1)}^{,ij}\beta^{(1)\prime}_{,ij}\;.
\end{eqnarray}

\subsection{Poisson gauge results}

Using the metric perturbations in the synchronous gauge presented in Section 2, 
we can now fix the parameters of the first-order gauge transformation as 
\begin{eqnarray}
\alpha^{(1)}&=&\frac{\tau}{3} \varphi \;, \\
\beta^{(1)}&=&\frac{\tau^2}{6} \varphi \;,
\end{eqnarray}
and $d^{(1)i}=0$, because of the absence of linear vector modes. 
We can then compute the first-order metric perturbations in the Poisson gauge,
\begin{eqnarray}
\psi_P^{(1)}&=&\phi_P^{(1)}=\varphi \;, \\
\chi_{P ij}^{(1)}&=&\chi_{ij}^{\top(1)} \;.
\end{eqnarray}
These equations show the well-known result for scalar perturbations
in the longitudinal gauge and the gauge invariance for tensor
modes at the linear level. 

The parameters of the second-order transformation read \cite{mm97}
\begin{eqnarray}
\alpha^{(2)}&=&-\frac{2}{21} \tau^3 \Psi_0+\tau
\left(\frac{10}{9} \varphi^2 +4 \Theta_0\right) \;,
\\
\beta^{(2)}&=&\tau^4\left(\frac{1}{72}\varphi^{,i}\varphi_{,i}
-\frac{1}{42}\Psi_0\right)+\frac{\tau^2}{3}\left(\frac{7}{2}
\varphi^2+6\Theta_0\right) \;,
\end{eqnarray}
where
\begin{equation}
\nabla^2 \Theta_0= \Psi_0-\frac{1}{3}\varphi^{,i}
\varphi_{,i}
\end{equation}
and
\begin{equation}
\nabla^2 d^{(2)}_j=\tau^2\left(-\frac{4}{3}\varphi_{,j}
\nabla^2\varphi+\frac{4}{3}\varphi^{,i}\varphi_{,ij}
-\frac{8}{3}\Psi_{0,j}\right) \;.
\end{equation}
For the second-order metric perturbations, one obtains \cite{mm97}
\begin{eqnarray}
\psi_P^{(2)}&=&\tau^2\left(\frac{1}{6}\varphi^{,i}\varphi_{,i}
-\frac{10}{21}\Psi_0\right)+\frac{16}{3} \varphi^2+12  \Theta_0 \;,
\\
\phi_P^{(2)}&=&\tau^2\left(\frac{1}{6}\varphi^{,i}\varphi_{,i}
-\frac{10}{21}\Psi_0\right)+\frac{4}{3} \varphi^2-8  \Theta_0,
\\
\nabla^2 \omega_P^{(2)i}&=&-\frac{8}{3}\tau\left(\varphi^{,i}
\nabla^2\varphi-\varphi^{,ij}\varphi_{,j}+2\Psi_0^{,i}\right)\;,
\\
\chi_{P ij}^{(2)}&=&\tilde{\pi}_{ij} \;. 
\label{pg2}
\end{eqnarray}

Quite interesting is to write down the equation governing the evolution of the 
second-order tensor modes, $\chi_{Pij}^{(2)} = \tilde{\pi}_{ij}$, in the 
Poisson gauge. It reads
\begin{equation}
\tilde{\pi}_{ij}''+\frac{4}{\tau}\tilde{\pi}_{ij}'-\nabla^2 \tilde{\pi}_{ij}=
- \frac{40}{3} {\cal T}_{ij} \;,
\end{equation}
with the source ${\cal T}_{ij}$ defined in Section 3.2 [after Eq. (56)]. 

Note also that the expressions for $\psi_P^{(2)}$ and $\phi_P^{(2)}$ 
could have been recovered, except for the sub-leading time-independent terms,
by taking the weak-field limit of Einstein's theory (e.g. ref.
\cite{pjep93}) and then expanding in powers of the perturbation
amplitude. 
  
\section{Discussion}

As we have seen, both in the synchronous and in the Poisson gauge, second-order 
tensor modes are dynamically generated by the gravitational instability of
scalar fluctuations. The form of these modes is however quite different
in the two gauges. The synchronous-gauge tensor modes contain
four terms: the first one, $\propto \tau^4$, can be easily seen to 
represent a Newtonian contribution, describing the dynamical tidal 
induction acting from the environment on the fluid element, then there is 
a post-Newtonian term, $\propto \tau^2$, a constant 
post-post-Newtonian 
term, required by the vanishing initial conditions, having no obvious 
observational effects, and, finally, a wave-like piece, which has
just the usual form as free cosmological gravitational waves. 
Quite interesting is the fact that the Newtonian and post-Newtonian terms are 
dropped by the transformation leading to  the Poisson gauge, so making the 
physical meaning of our second-order tensor modes more transparent. 

It has been argued \cite{mt96} that a possible observational evidence 
for these tensor modes could be in terms of a tensor generalization of 
the so-called Rees-Sciama effect 
\cite{rs68,ma90,ma92,ma94,ar94,tu95,se96a,sa96}, 
a secondary anisotropy of the Cosmic Microwave Background. 
Let us discuss this point in more detail. 
Mollerach \& Matarrese \cite{mm97} have recently obtained a general formula 
for the full (scalar, vector and tensor) Rees-Sciama effect. Their 
expression reads
\begin{equation}
\delta T_{RS}=\frac{1}{2}\int_{\lambda_{\cal{O}}}^{\lambda_{\cal{E}}}
d\lambda \left(\psi^{(2)'}+\phi^{(2)'}+\omega^{(2)'}_i e^i-
\frac{1}{2}\chi^{(2)'}_{ij}e^i e^j\right) \;, 
\end{equation}
where $e^i$ represents the unit vector of the unperturbed incoming photon
and the integration is along the unperturbed photon path. The first two terms 
correspond to the scalar contribution to the effect, while the third
and fourth ones to the vector and tensor contributions, respectively. 
This splitting of the contributions depends strongly on the gauge choice.
For example, it can be shown that in the Poisson gauge the contributions 
coming from the vector and tensor pieces are subdominant with respect to 
the scalar ones, while in the synchronous gauge the scalar and tensor 
contributions are of the same order of magnitude.

One may then wonder whether there is any hope to detect the cosmological 
stochastic gravitational-wave background produced by second-order scalar 
fluctuations, which we have discussed in these notes. The general 
prospects for detecting a stochastic gravitational-wave background have been 
nicely reviewed by Bruce Allen at this meeting (see also Ref. \cite{allen}). 
It is, of course, the oscillating part of $\chi^{(2)\top}_{ij}$ which is of 
relevance for earth or space detectors. The problem for these wave-like modes 
is that their energy density suffers the usual $a^{-4}$ dilution 
caused by free-streaming inside the Hubble radius, while on the 
horizon scale their closure density is already extremely small, 
$\Omega_{gw} \sim \delta_H^4$ (where $\delta_H$ is the {\em rms} density 
contrast at horizon-crossing), because of their secondary origin. 
More promising is the possibility that a non-negligible amount of 
gravitational radiation is produced during the strongly non-linear 
stages of the collapse of cosmic proto-structures. 
This issue would however require a non-perturbative approach, which is 
beyond the available analytical techniques in this field. 
A first step in the direction of estimating the amplitude of such an effect has 
been recently made by Matarrese \& Terranova \cite{mt96}, who studied the 
collapse of homogeneous triaxial ellipsoids embedded in an 
Einstein-de Sitter universe, including post-Newtonian tensor modes, and
showed that these post-Newtonian terms tend to dominate over
the Newtonian ones, during the late stages of collapse.

\acknowledgements

This work has been partially supported by the Italian MURST and by 
the Vicerrectorado de investigaci\'on de la Universidad de Valencia.
We would like to thank M. Bruni and S. Sonego for many useful discussions.

\vfill
\end{document}